\documentclass[sigconf] {acmart}

\usepackage{hyperref}
\usepackage{amsmath,amsfonts}
\usepackage{amsthm}
\usepackage{algorithmic}
\usepackage{graphicx}
\usepackage{textcomp}
\usepackage{xcolor}
\usepackage{caption}
\usepackage{tikz}
\usepackage{float}
\usepackage{grumble}
\usepackage{comment}
\usepackage{subcaption}
\usepackage{enumitem}
\usepackage{listings}
\usepackage{longtable}
\usepackage{url}
\usepackage{fancyhdr}
\usepackage{booktabs}
\usepackage{tagging}
\usepackage{color}
\usepackage{colortbl}

\copyrightyear{2025}
\acmYear{2025}
\setcopyright{cc}
\setcctype{by}
\acmConference[SC '25]{The International Conference for High Performance Computing, Networking, Storage and Analysis}{November 16--21, 2025}{St Louis, MO, USA}
\acmBooktitle{The International Conference for High Performance Computing, Networking, Storage and Analysis (SC '25), November 16--21, 2025, St Louis, MO, USA}
\acmDOI{10.1145/3712285.3759785}
\acmISBN{979-8-4007-1466-5/2025/11}

\definecolor{codegreen}{rgb}{0,0.6,0}
\definecolor{codegray}{rgb}{0.5,0.5,0.5}
\definecolor{codepurple}{rgb}{0.58,0,0.82}
\definecolor{backcolour}{rgb}{0.95,0.95,0.92}
\definecolor{customgreen}{rgb}{0,0.5,0}

\lstdefinestyle{qonductorYAML}{
    backgroundcolor=\color{backcolour},   
    commentstyle=\color{codegreen},
    keywordstyle=\color{customgreen},
    numberstyle=\tiny\color{codegray},
    stringstyle=\color{codepurple},
    basicstyle=\ttfamily\scriptsize,
    breakatwhitespace=false,         
    breaklines=true,                 
    captionpos=b,                    
    keepspaces=true,                 
    numbers=left,                    
    numbersep=5pt,                  
    showspaces=false,                
    showstringspaces=false,
    showtabs=false,                  
    tabsize=2,
    keywords={spec, containers, name, image, resources, limits, qubits, nvidia, com, gpu, qpu, quantum, ibm}
}

\lstdefinestyle{qonductorListing}{
    backgroundcolor=\color{backcolour},   
    commentstyle=\color{codegreen},
    keywordstyle=\color{magenta},
    numberstyle=\tiny\color{codegray},
    stringstyle=\color{codepurple},
    basicstyle=\ttfamily\scriptsize,
    breakatwhitespace=false,         
    breaklines=true,                 
    captionpos=b,                    
    keepspaces=true,                 
    numbers=left,                    
    numbersep=5pt,                  
    showspaces=false,                
    showstringspaces=false,
    showtabs=false,                  
    tabsize=2,
    keywords={from, import, }
}

\newcommand{\myparagraph}[1]{\smallskip \noindent{\bf {#1}.}}
\newcommand{\projecttitle}{Qonductor}

\setlist{noitemsep,topsep=0pt,parsep=0pt,partopsep=0pt}

\usepackage[subtle]{savetrees}

\usepackage{setspace}
\usepackage[margin=5pt,font={stretch=0.9}]{caption}

\begin{document}
\pagestyle{empty}   

\title{\projecttitle{}: A Cloud Orchestrator for Quantum Computing}


\author{Emmanouil Giortamis}
\affiliation{%
  \institution{Technical University of Munich}
  \city{Munich}
  \country{Germany}
}
\email{emmanouil.giortamis@tum.de}
\orcid{0009-0000-3638-2969}

\author{Francisco Romão}
\affiliation{%
  \institution{Technical University of Munich}
  \city{Munich}
  \country{Germany}
}
\email{francisco.romao@tum.de}
\orcid{0009-0004-0145-9785}

\author{Nathaniel Tornow}
\affiliation{%
  \institution{Technical University of Munich}
  \city{Munich}
  \country{Germany}
}
\email{nathaniel.tornow@tum.de}
\orcid{0009-0000-2095-2242}

\author{Dmitry Lugovoy}
\affiliation{%
  \institution{Technical University of Munich}
  \city{Munich}
  \country{Germany}
}
\email{dmit.lugovoy@tum.de}
\orcid{0009-0008-0244-0225}

\author{Pramod Bhatotia}
\affiliation{%
  \institution{Technical University of Munich}
  \city{Munich}
  \country{Germany}
}
\email{pramod.bhatotia@tum.de}
\orcid{0000-0002-3220-5735}



\pagestyle{empty}

\begin{abstract}


We describe \projecttitle{}, a cloud orchestrator for hybrid quantum-classical applications that run on heterogeneous hybrid resources. \projecttitle{}
abstracts away the complexity of hybrid programming and resource management by exposing the \textit{\projecttitle{} API}, a high-level and hardware-agnostic API. The \textit{resource estimator} strategically balances quantum and classical resources to mitigate resource contention and the effects of hardware noise.  
The \textit{hybrid scheduler} automates job scheduling on hybrid resources and balances the tradeoff between users' objectives of QoS and the cloud operator's objective of resource efficiency.

We implement an open-source prototype and evaluate \projecttitle{} using more than 7000 real quantum runs on the IBM quantum cloud to simulate real cloud workloads. \projecttitle{} achieves up to 54\% lower job completion times (JCTs) while sacrificing 3\% execution quality, balances the load across QPU, which increases quantum resource utilization by up to 66\%, and scales with growing system sizes and loads. 

\end{abstract}

\maketitle 
\thispagestyle{empty}

\section{Introduction}
\label{sec:introduction}

Quantum computing offers the potential to solve computational problems beyond the capabilities of classical computers by leveraging the principles of quantum mechanics \cite{arute2019quantum, daley2022practical, grover1996fast, shor1999polynomial}. Quantum computing is realized in the form of Quantum Processing Units (QPUs) \cite{chi2022programmable}, which are characterized by inherent noise and small qubit counts \cite{preskill2018quantum} and are now offered by all major cloud providers in a quantum-as-a-service fashion \cite{ibmQuantum, aws-quantum, google-quantum, azure-quantum}.


However, since QPUs are not general-purpose processors, the quantum programming and execution models are \textit{hybrid}, consisting of classical and quantum code. For instance, quantum applications can use classical pre- and post-processing steps to mitigate or correct hardware noise errors \cite{tannu2019mitigating, tornow2025qvm, das2021jigsaw}. These steps often leverage classical accelerators such as GPUs \cite{tang2022scaleqc} or FPGAs \cite{maurya2023scaling} for improved performance.

On top of that, the quantum cloud landscape is characterized by resource contention. Specifically, due to manufacturing and operational requirements \cite{krinner2019engineering}, QPUs are an expensive and scarce resource, with less than 100 QPUs available \textit{globally} by all cloud vendors combined \cite{ibmQuantum, aws-quantum, google-quantum, azure-quantum}, with the largest provider, IBM, typically offering fewer than ten online at any given time \cite{ibmQuantum}. In contrast, demand is constantly increasing, with IBM recently celebrating \textit{three trillion} program executions on their platform \cite{qiskit-trillion-circuits}.

Despite this scarcity, QPUs are vastly heterogeneous since there is an abundance of quantum technologies, architectures, and models, each presenting different tradeoffs between performance metrics, manufacturing complexity, and operational requirements \cite{gyongyosi2019a}. More importantly, heterogeneity extends to QPU performance, with same-model QPUs experiencing significant performance differences (we detail this in \S\ref{section:motivation:characteristics-nisq}), and this performance changes over time unpredictably \cite{ravi2022quantum, patel2020experimental}.

\begin{figure*} [ht]
    \centering
    \includegraphics[width=0.9\textwidth]{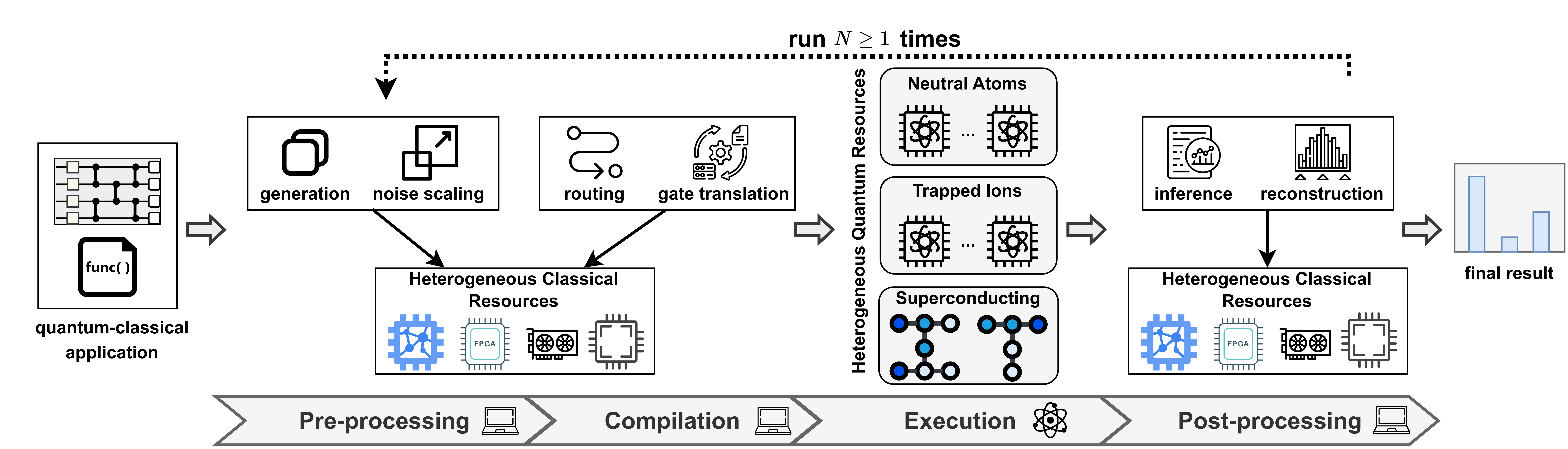}
    \caption{Quantum cloud and hybrid computational model (\S~\ref{section:background:quantum-cloud}). {\em Quantum applications are hybrid, i.e., require quantum and classical resources. The pre-processing, compilation, and post-processing steps run on classical heterogeneous accelerators. QPUs are vastly heterogeneous across space and time, i.e., QPU technologies, architectures, and calibration data. }}
    \label{fig:quantum-cloud}
    \vspace{-3pt}
\end{figure*}

Naturally, QPU heterogeneity and scarcity drive users to select the highest-fidelity QPUs available, inevitably leading to QPU load imbalance, where best-performing QPUs become hotspots while the rest are underutilized \cite{ravi2021adaptive, ravi2022quantum, wang2024qoncord}. Consequently, there is an inherent conflict between achieving high fidelity and maintaining low job completion times (JCTs), as users ideally desire both but must often compromise on one, typically sacrificing JCTs. 

To summarize, quantum application development and orchestration are characterized by hybrid workflows and resources, scarce and vastly heterogeneous QPUs, and  fundamentally conflicting objectives, posing three critical challenges.

\textbf{First, hybrid programming and execution models are required}. The standard practice for developing hybrid applications is through tedious and \textit{manual} composition of classical and quantum tasks into workflows with virtually no standardization. Users navigate a largely heterogeneous landscape unguided to \textit{manually} select the resources required to execute their workflows, amplifying QPU load imbalance \cite{wang2024qoncord, ravi2022quantum, giortamis2025qos}.
\textbf{Second, quantum performance estimation depends on classical resources}. While research and industry efforts are constantly developing classical error mitigation techniques \cite{das2021jigsaw, tannu2019mitigating, tornow2025qvm}, their impact on quantum performance metrics (i.e., fidelity and execution time) is typically not included in the cloud scheduling decisions \cite{ravi2021adaptive, wang2024qoncord, stein2022eqc}. 
\textbf{Third, the conflicting objectives of the quantum cloud require multi-objective optimization}. Designing a system for the growing cloud that doesn't operate at the extremes of the fidelity-JCT tradeoff, i.e., does not simply choose the least busy or the highest-fidelity machine \cite{ravi2021adaptive, ibm-qiskit-least-busy}, requires scalable multi-objective scheduling algorithms. 



Thus, we pose the following research question: \textit{How to design a scalable hybrid quantum-classical orchestrator that balances the conflicting objectives of the hybrid cloud?}

To answer this question, we introduce \projecttitle{}, a cloud orchestrator for deploying hybrid applications on hybrid and heterogeneous clusters. \textbf{First,} the \textit{\projecttitle{} APIs} abstract away the complexity of hybrid application development and execution using hybrid resources. \textbf{Second,} our \textit{resource estimator} systematically explores hybrid resource configurations that reduce resource contention while increasing execution quality (or fidelity). \textbf{Third,} our \textit{hybrid scheduler} balances the tradeoff between fidelity vs. job completion times (JCTs).

We implement an open-source prototype of \projecttitle{} in Python \cite{python} and Go \cite{go-lang} by building on top of Kubernetes'  scheduler, key-value store, and custom resource definitions to support QPUs \cite{kubernetes}, and the resource estimator based on the Qiskit framework \cite{qiskit}, the scikit-learn ML library \cite{scikit-learn}, and the pymoo optimization library \cite{pymoo}.

To evaluate \projecttitle{}'s effectiveness, we first analyze real quantum cloud load conditions, construct a simulation environment resembling this workload, and evaluate \projecttitle{} in this environment using more than 70.000 benchmark circuits. Our results show that \projecttitle{}: (1) achieves up to 54\% lower JCTs for a $\sim 3$\% fidelity penalty on average, (2) evenly balances the load across QPUs and achieves $66\%$ higher QPU utilization, (3) accurately estimates fidelities and runtimes in at least $\sim75\%$ of the times, (4) scales linearly with an increasing cluster size and up to 3$\times$ the current quantum cloud load.

\myparagraph{Contributions} We make the following contributions:

\begin{enumerate}[leftmargin=*]

    \item \textbf{Exposing hybrid cloud tradeoffs:}
    We demonstrate that quantum performance can be increased using much cheaper classical resources and that users can experience vastly lower waiting times for minimal fidelity penalties.  

    \item \textbf{Hardware-agnostic programming model:} We introduce a hardware-agnostic API that simplifies programming hybrid applications and abstracts the underlying heterogeneous resources away.

    \item \textbf{Hybrid resource estimation:} We introduce hybrid quantum-classical resource estimation, the first systematic hardware-aware estimation of fidelity, runtime, and cost (\$) when involving heterogeneous hybrid resources.      

    \item \textbf{Hybrid scheduler:} We propose the first hybrid scheduler that balances the tradeoff between the conflicting objectives of fidelity vs. JCTs by employing Pareto-optimal multi-objective optimization techniques.

\end{enumerate}

\section{Background}
\label{section:background}


\subsection{Quantum Computing Basics}
\label{section:background:101}

\myparagraph{Noisy quantum hardware}
Modern quantum hardware, classified as noisy intermediate-scale quantum (NISQ) devices \cite{preskill2018quantum}, operates with tens to a few hundred qubits \cite{ibmQuantum} and is affected by a variety of error channels. Quantum operations deviate from their ideal unitary evolution due to stochastic Pauli errors, decoherence-induced amplitude damping and phase damping, and control inaccuracies that lower gate fidelities \cite{google-nisq-properties}. Moreover, idle qubits experience state degradation through T1 relaxation and T2 dephasing \cite{klimov2018fluctiations}, while unwanted qubit-qubit interactions induce correlated errors via crosstalk \cite{cross2019validating}. The probabilities of these errors are characterized during periodic calibration procedures \cite{tornow2022minimum}, which yield comprehensive datasets—detailing parameters such as T1, T2, and gate fidelities—that are publicly available \cite{google-calibration, ibmQuantum} but can fluctuate unpredictably between calibration cycles.

\begin{figure*}[ht]
  \centering
  \includegraphics[width=\textwidth]{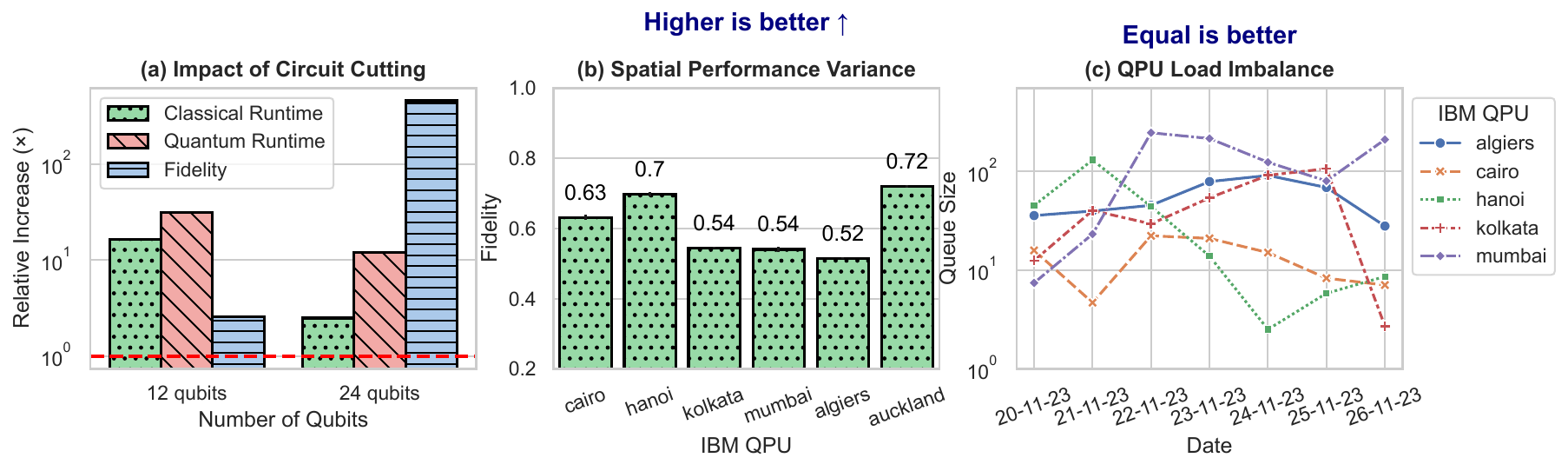}
  \caption{Quantum orchestration challenges (\S~\ref{section:motivation:characteristics-nisq}). {\bf (a)} {\em  Impact of circuit cutting as a relative increase in execution fidelity, quantum, and classical runtime for 12-qubit and 24-qubit circuits. } {\bf (b)} {\em  Spatial performance variance: fidelity of a 12-qubit GHZ circuit on different IBM QPUs. There is a 38\% fidelity difference from best to worst QPU.} {\bf (c)} { 
QPU load imbalance: number of pending jobs on different IBM QPUs. There is up to $\sim$100$\times$ load difference across QPUs.}}
  \label{fig:nisq-cloud-characterization}
\end{figure*}

\myparagraph{Quantum performance metric}
To evaluate the quality of circuit execution on NISQ devices, we use the \textit{Hellinger fidelity} metric \cite{fidelity-qiskit}, which quantifies the similarity between the noisy probability distribution obtained from the actual device and the ideal distribution that would be produced by noiseless, perfect hardware. Fidelity ranges from 0 to 1, with higher values indicating better quality results.

\myparagraph{Quantum error mitigation}
A suite of techniques has been developed to reduce the impact of noise on quantum computations without the need for full fault tolerance. These methods generally adhere to a three-stage workflow: (1) optimization and/or generation of  circuit(s), (2) execution on noisy quantum hardware, and (3) post-processing to reconstruct a less noisy result.  Among these methods, Zero-Noise Extrapolation (ZNE) infers the zero-noise limit by running circuits at different noise levels \cite{li2017efficient, temme2017error}; Probabilistic Error Cancellation (PEC) models noise inverses via probabilistic resampling \cite{temme2017error, endo2018practical}; Dynamical Decoupling (DD) applies pulse sequences to suppress decoherence \cite{pokharel2018demonstration, das2021adapt}; Readout Error Mitigation (REM) corrects measurement errors \cite{bravyi2021mitigating}, and Pauli Twirling converts general noise into stochastic Pauli noise for easier correction \cite{wallman2016noise}.

\subsection{Quantum Cloud Computing}
\label{section:background:quantum-cloud}

\myparagraph{Quantum cloud and scarcity of quantum resources}
Major cloud providers such as IBM, Microsoft Azure, Google Cloud, and AWS currently offer access to quantum processing units (QPUs) \cite{ibmQuantum, azure-quantum, google-quantum, aws-quantum}. Globally, access is offered to fewer than $100$ QPUs, while IBM alone recently celebrated three \textit{three trillion} circuit executions on their cloud \cite{qiskit-trillion-circuits}. Notably, the gap between the growing demand for quantum resources and the current supply cannot be closed by building more QPUs, which are expensive to build and operate. 



\myparagraph{Hybrid computational model}
In practice, quantum applications are inherently hybrid, relying on both quantum and classical computing. This is because QPUs are not standalone processors: classical computers are needed to compile quantum programs \cite{qiskit-transpiler, kirmenis2025weaver} and to handle noise mitigation or error correction (\S~\ref{section:background:101}). The typical workflow and resources required for a quantum application are shown in Figure \ref{fig:quantum-cloud}. First, the quantum circuit is pre-processed to prepare error mitigation techniques. %
Then, the circuit is compiled to a target QPU to match the QPU's constraints, e.g., basis gate set, and the circuit is executed on one or more QPUs. 
Lastly, the execution results are typically post-processed to reconstruct the less noisy result, typically through inference \cite{temme2017error, li2017efficient}.




\myparagraph{Heterogeneous hybrid cloud resources}
As shown in Figure \ref{fig:quantum-cloud}, the classical processes include CPUs and specialized accelerators (xPUs, FPGAs, etc.). For instance, GPUs and Tensor Processing Units (TPUs) can be used for circuit knitting \cite{tang2022scaleqc, tornow2024quantum}, while FPGAs are used for qubit readout classification \cite{maurya2023scaling}. 

At the same time, the quantum cluster is heterogeneous in three dimensions: \textbf{(1)} There exist multiple QPU technologies, such as superconducting \cite{arute2019quantum}, trapped ions \cite{cirac1995quantum}, and neutral atoms \cite{henriet2020quantum}. These technologies involve trade-offs between performance metrics, manufacturing complexity, and operational requirements \cite{gyongyosi2019a}. 
\textbf{(2)} Different architectures of same-technology QPUs vary in qubit topologies, basis gate sets, and noise models. \textbf{(3)} In fact, QPUs even of the \textit{same} model have different noise models, which vary across calibration cycles, leading to spatiotemporal performance variance \cite{ravi2022quantum, patel2020experimental, giortamis2025qos}, as we detail in \S~\ref{section:motivation}.

\section{Why is Hybrid Orchestration Challenging?}
\label{section:motivation}

\label{section:motivation:characteristics-nisq}

Managing the quantum cloud faces distinct challenges compared to classical orchestration and resource management: primitive programming and execution models, lack of hybrid resource estimation, and conflicting optimization objectives.

\myparagraph{\#1: Primitive programming and execution models}
Developing and running quantum applications today involves hybrid quantum-classical code and resources (\S~\ref{section:background:quantum-cloud}). 
Unfortunately, the prevailing programming model remains primitive: developers must \textit{manually} stitch together classical and quantum logic—typically in Python—and \textit{explicitly} manage low-level execution details. This includes selecting quantum devices, configuring classical control logic, and managing execution workflows. Such manual composition introduces unnecessary complexity, increases the chance of user error, and typically is tightly coupled to specific backends.



{\bf \underline{Key idea \#1:}}  We need hardware-agnostic APIs to enable transparent development and execution of hybrid workflows on heterogeneous hybrid resources.


\begin{table}[t]
\caption{IBM Cloud Pricing.}
    \centering
        \begin{tabular}{|c|c|c|}
            \hline
            Resource Type & Price/Task & Price/Hour \\
            \hline \hline
            Standard VM & $<1$\$ & 1-5\$ \\
            \hline
            High-end VM & 1-10\$ & 10-40\$ \\
            \hline
            QPU & 30-200\$ & 3000-6000\$ \\
            \hline
        \end{tabular}
    \label{tab:cloud_costs}
\end{table}

\myparagraph{\#2: Hybrid resource estimation}
Quantum execution fidelity is closely tied to classical compilation and pre- and post-processing steps, which often introduce additional runtime overhead to improve fidelity. This is shown in Figure \ref{fig:nisq-cloud-characterization} (a), where we use the circuit knitting error mitigation technique \cite{mitarai2021constructing, tornow2025qvm} to cut 12-qubit and 24-qubit circuits in half and execute them sequentially on the same QPU. In the 24-qubit case, although the average classical and quantum runtimes increase by $2.5\times$ and 12$\times$, respectively, the average fidelity increases by $\sim 450 \times$. 

Notably, classical resources are significantly cheaper and more accessible; e.g., while the IBM Cloud offers less than $20$ QPUs, it offers thousands of classical servers \cite{ibm-classical}. Table \ref{tab:cloud_costs} shows the price (in \$) per classical/quantum task/hour for different resource types. Standard VMs comprise 4-32 vCPUs and 16-64 GB RAM, while high-end VMs comprise 64+ vCPUs and up to 6 TB RAM. Notably, even the high-end VM-hours cost two orders of magnitude less than QPU-hours. 

{\bf \underline{Key idea \#2:}} We can increase quantum execution fidelity by leveraging error mitigation techniques (\S~\ref{section:background:101}), which require using the widely cheaper and more abundant classical resources.


\myparagraph{\#3: Quantum cloud design trade-offs}
The quantum cloud is characterized by conflicting objectives between the users' Quality-of-Service requirements (high fidelity and low JCTs) and the cloud operator's requirements (resource efficiency), caused by QPU spatiotemporal heterogeneity and the scarcity of QPUs.

First, QPU noise characteristics differ significantly across space and time, in contrast to classical processors. Specifically, execution fidelity can fluctuate across different QPUs and different calibration cycles \cite{giortamis2025qos, ravi2022quantum, wang2024qoncord, smith2023fast} as shown in Figure \ref{fig:nisq-cloud-characterization} (b), where we run a 12-qubit GHZ circuit on six IBM 27-qubit QPUs on 08-11-23. Fidelity varies across them, with up to 38\% higher fidelity in \emph{auckland} than \emph{algiers}.

This performance variance naturally motivates users to select the highest-fidelity QPUs. Figure \ref{fig:nisq-cloud-characterization} (c) shows the number of pending jobs for every QPU and every day of a week in November 2023. QPUs face up to two orders of magnitude load difference, e.g., on 26-11-23, \emph{mumbai} faces $\sim$100$\times$ more pending jobs than \emph{kolkata}.

Evidently, there is an inherent fidelity-JCT tradeoff. Ideally, users want the fidelity of the \textit{highest-fidelity} QPU with the waiting time of the \textit{least-busy} QPU. However, to maximize fidelity, all incoming jobs must be scheduled on the highest-fidelity QPU(s), forming hotspots, increasing the average JCTs, and decreasing average QPU utilization. Conversely, to minimize JCTs (and increase utilization), the jobs must be evenly distributed across all QPUs, decreasing average fidelity.


{\bf \underline{Key idea \#3:}}  We trade \textit{minimal} fidelity penalties for \textit{significant} JCT reduction. Combined with error mitigation (key idea \#2), the fidelity loss can even be compensated. 

\section{Overview}
\label{section:overview}


We propose \projecttitle{}, a scalable cloud orchestrator for developing and deploying hybrid applications on heterogeneous resources. Our system design is based on the key ideas presented in \S~\ref{section:motivation:characteristics-nisq} to address the challenges of quantum orchestration. The system comprises the \textit{data plane}, used to deploy, invoke, and store hybrid workflow images; the \textit{control plane}, which manages worker nodes and performs hybrid resource estimation and scheduling; \textit{worker nodes} that manage the underlying classical accelerators and QPUs, and the \textit{system monitor} that persists \projecttitle{}'s state. 

\subsection{\projecttitle{} Architecture}
\label{section:overview:components}

\begin{figure}[t]
    \centering
    \includegraphics[width=\columnwidth]{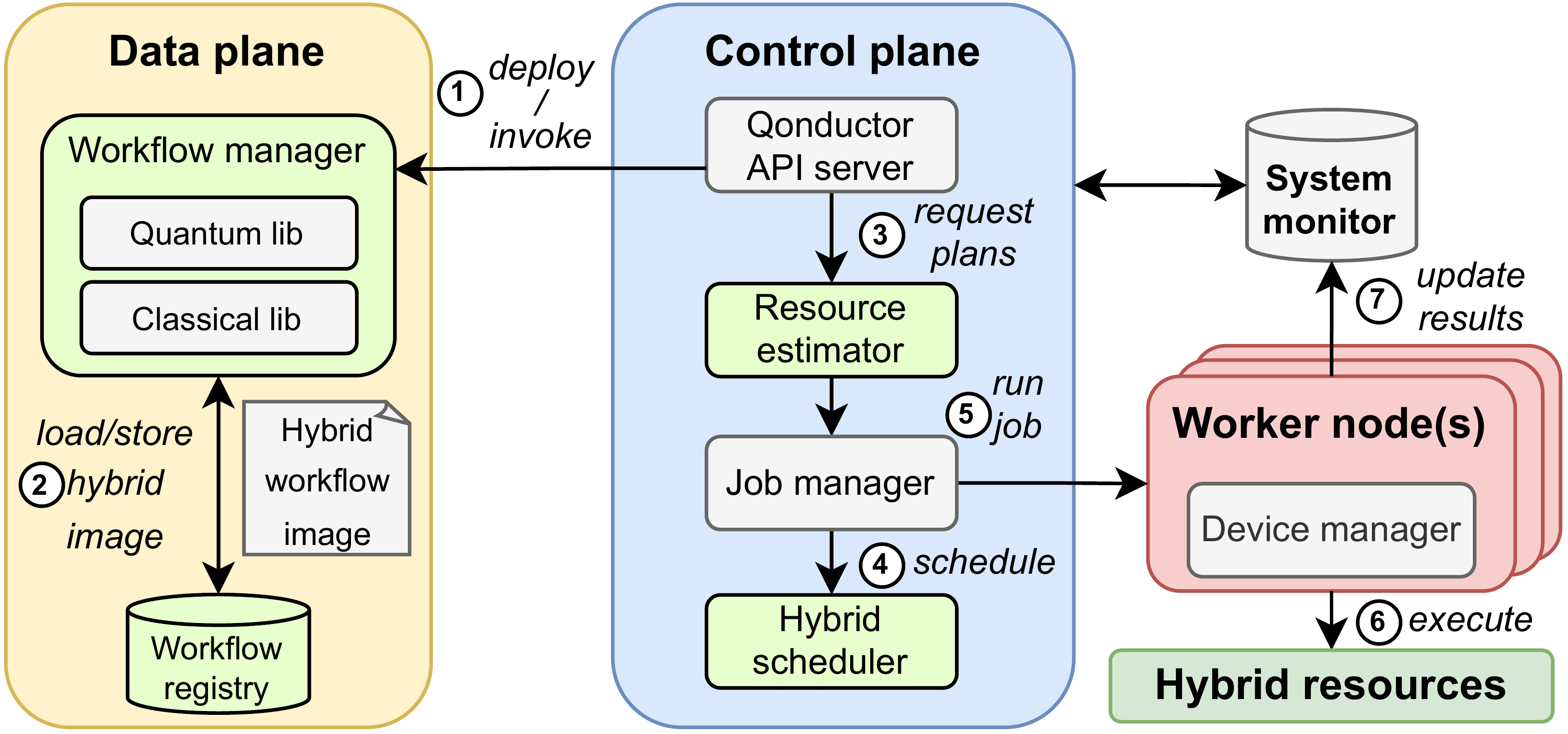}
    \caption{\projecttitle{} overview (\S~\ref{section:overview}). {\em Qonductor comprises the control plane, data plane, worker node(s), and the system monitor. Core components are highlighted as light green boxes.
    The control plane performs resource estimation, job management, and hybrid scheduling. The data plane is used to deploy and invoke hybrid images. Workers manage hybrid resources.}} 
    \label{fig:system_overview}
\end{figure}

We detail the architecture and core components of \projecttitle{}, as shown at a high level in Figure \ref{fig:system_overview}.

\myparagraph{Data plane}
The data plane provides functionality for configurable, programmable, and reusable hybrid application development and execution. To achieve this, we implement the \textit{workflow manager} that offers libraries of commonly used quantum algorithms (e.g., QAOA \cite{farhi2014quantum}) and classical functions (e.g., error mitigation \cite{mitiq}). To minimize repetitive and manual hybrid application development, the workflow manager packages hybrid applications and user configuration (e.g., accelerator or QPU preferences) into \textit{hybrid workflow images} that are persisted in the \textit{workflow registry}. This enables users to reuse existing hybrid workflows out of the box with minimal effort and distribute them. We elaborate on the \projecttitle{} programming model and the data plane's components in \S~\ref{section:programming-model}.

\myparagraph{Control plane}
The control plane is the core component of \projecttitle{} and is responsible for managing hybrid workflow execution while achieving resource efficiency and improving quality of service by leveraging hybrid resource estimation and scheduling. In more detail, the \textit{\projecttitle{} API server} is the interface between the users and \projecttitle{}. When users invoke hybrid workflow images, the API server calls the \textit{resource estimator} to request resource plans for the users. Given the generated resource plan(s), the \textit{job manager} iterates over the workflow's jobs and, for each job, requests a resource allocation from the \textit{hybrid scheduler}. The scheduler allocates resources while balancing fidelity and JCTs, guided by the resource plan and user's execution configuration. Lastly, the job manager runs each workflow job on the worker node(s) assigned by the scheduler. We elaborate on the resource estimator in \S~\ref{section:resource_estimator} and the hybrid scheduler in \S~\ref{section:scheduler}.

\myparagraph{Worker nodes}
 The worker nodes serve two roles: (1) Execute jobs on their underlying devices (classical accelerators or QPUs) and (2) monitor and update the device status. For the first role, the \textit{device manager} spawns containers that run the job on the node. For the second, the device manager periodically queries the classical nodes and QPUs to get static (e.g., number of cores/qubits) and the current dynamic information (e.g., queue sizes, utilization, calibration data), and updates the system monitor accordingly. For the QPU calibration data specifically, it fetches the new calibration data after each calibration cycle and updates the system monitor accordingly.

\myparagraph{System monitor} The system monitor is a datastore where the complete system state is persisted. Specifically, the datastore maintains the list of available worker nodes and their resources, i.e., the number of cores, memory, accelerators, etc. (static information), and their current utilization, job queues, live status, etc. (dynamic information). Specifically for QPUs, we store the QPUs' architectures, coupling maps, number of qubits, etc. (static information), and the current job queues and calibration data (dynamic information). The datastore also stores workflow information, specifically execution status (e.g., failed, completed, running, etc.), resource allocations, and their intermediate or final results. 

\myparagraph{Fault tolerance}
\label{section:architecture:fault-tolerance}
The system relies on the control plane and the system monitor; therefore, it is crucial to make both fault-tolerant. We use a quorum of $2f+1$ nodes to replicate the components of the control plane, with $f=1$ by default. The backup replicas detect the component's failures through heartbeat messages that experience delays greater than $\Delta$, since we assume a partially synchronous message model \cite{dwork1988consensus}. In case of failure, the backups elect a new leader using Raft \cite{ongaro2014search}. The same setup applies to the system monitor datastore. 


\begin{table}[t]
\fontsize{8}{9}\selectfont 
\caption{\projecttitle{} programming API. CP and DP stand for control and data plane, respectively.} 

\begin{center}
\begin{tabular}{ |c|c|c| }
 \hline
 \bf{Operation} & {\bf Caller} & {\bf Callee} \\ \hline
 Create a workflow with hybrid code. & User &  CP \\
 Deploy a workflow. & User & CP \\
 Invoke a workflow. & User & CP \\
Get the workflow results. & User & CP \\
 Register a workflow image in the registry. & DP & DP \\
 List available hybrid workflow images. & CP &  DP\\
 Estimate the hybrid resources required. & CP & CP \\
 Generate a schedule for hybrid tasks. & CP & CP \\
 
  \hline

\end{tabular}
\end{center}
\vspace{-2pt}
\label{tab:api}

\end{table}

\subsection{System Workflow}
\label{section:overview:workflow}

The system workflow is shown in Figure \ref{fig:system_overview}. Users call \texttt{invoke/deploy} through the \projecttitle{} API server to deploy hybrid workflows or invoke them, respectively (1). The workflow manager loads/stores hybrid workflow images from the workflow registry, depending on the user's call (2). Then, the API server requests the resource estimator to generate resource plans that trade fidelity for runtime cost (3). The job manager invokes the scheduler to allocate worker nodes for the workflow's jobs  that comply with the resource plan and the user's preferences (4). Finally, the job manager runs the job on the selected worker nodes (5), which execute it (6) and update the execution results (7).

\section{\projecttitle{} Programming Model}
\label{section:programming-model}

We introduce the \projecttitle{} programming model designed to abstract away the complexity of programming and executing hybrid workflows. Clients can either reuse existing images from the workflow registry or create new ones with the help of (1) libraries of quantum and classical routines, (2) automated workflow generation and image packaging, and (3) hardware-agnostic deployment.

\myparagraph{Classical and quantum libraries}
The extensible libraries of commonly used quantum and classical functions aid programmability.
Specifically, the classical library contains error mitigation techniques \cite{qiskit-error-mitigation, mitiq, tornow2025qvm} and simulation libraries \cite{nvidia-quantum, pennylane}. The quantum library includes state-of-the-art quantum algorithms such as the Variational Quantum Eigensolver (VQE) \cite{peruzzo2014variational}, the Quantum Approximate Optimization Algorithm (QAOA) \cite{farhi2014quantum}, and the Quantum Fourier Transform (QFT) \cite{weinstein2001implementation}, among others. 

\myparagraph{Workflow image generation}
The workflow manager automatically splits a Python file into quantum and classical code files while maintaining library dependencies and keeping track of input/output data between the files. Then, the manager creates a directed acyclic graph (DAG) $G=(V,E)$ where $V$ is the set of classical and quantum steps and $E=\{(E_i,E_j) \in  V\times V \}$ are the control and data flow dependencies between them. The leader node's job manager later leverages this graph representation to handle workflow scheduling and execution. Lastly, the workflow graph model, the hybrid code files, and the execution configuration files are packed into a hybrid workflow image and stored in the workflow registry.

\myparagraph{Workflow registry}
Users typically write the same hybrid applications repeatedly, which becomes tedious for complex workflows. To streamline the deployment of such applications, the workflow registry is a repository for ready-to-execute workflow images. Users can leverage the registry to distribute or execute these images by providing input and customizing the execution to suit their unique requirements. Listing \ref{code:YAML} shows two example images (L4 and L9), one for error mitigation using CUDA and one for a QAOA algorithm.

\begin{lstlisting}[frame=h,style=qonductorYAML,
                caption=Example YAML deployment configuration file., label={code:YAML}]
spec:
  containers:
  - name: qaoa-error-mitigated
    image: nvidia/cuda:11.0-base
    resources:
      limits:
        nvidia.com/gpu: 1  # Request one GPU
  - name: qaoa-algorithm
    image: qaoa:latest
    resources:
      limits:
        quantum.ibm.com/qpu: 1  # Request one QPU
        qubits: 20 # Request QPU size >= 20
\end{lstlisting}

\begin{figure*} [t]
    \centering
    \includegraphics[width=\textwidth]{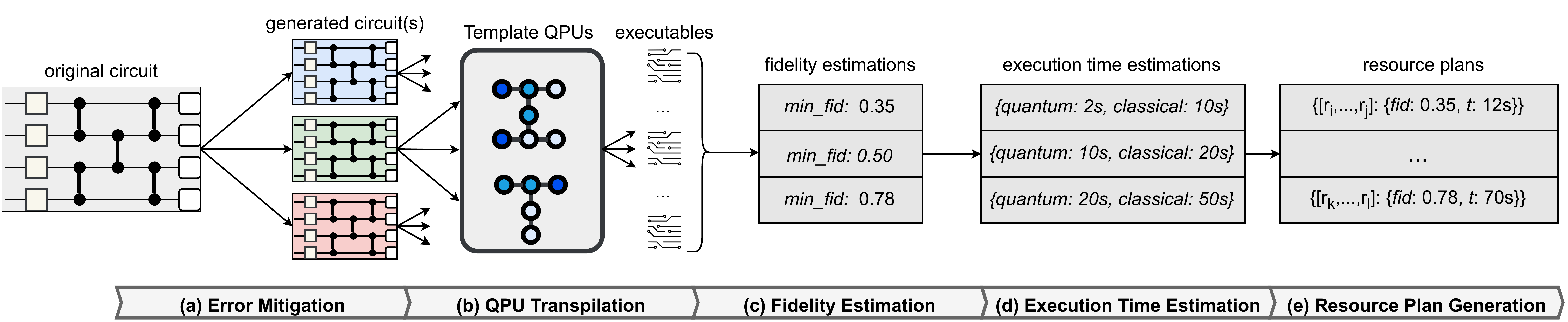}
    \caption{Resource estimator workflow (\S~\ref{section:resource_estimator}). {\em \textbf{(a)} Error mitigation is applied to the original circuit, which generates one or more circuits. \textbf{(b)} The generated circuits are compiled for template QPUs, generating executable circuits. \textbf{(c)} Fidelity is estimated for all generated circuits where the minimum fidelity binds the solution fidelity. \textbf{(d)} The classical and quantum task execution times are estimated. \textbf{(e)} Resource plans are generated based on the estimated fidelities and total execution times.}}
    \label{fig:resource-estimator}
    
\end{figure*}

\myparagraph{Hybrid execution configuration}
Users can customize computational resources in \projecttitle{} by requesting specific QPUs or classical accelerators. Listing \ref{code:YAML} shows an example YAML execution configuration file where the user requests at least one GPU (L7) and a QPU with at least 20 qubits (L12-L13).



\begin{lstlisting}[frame=h,style=qonductorListing,
                language={Python},
                caption= Example usage of the \projecttitle{} APIs., label={code:programming_model}]
from qonductor.lib.quantum import QAOA
from qonductor.lib.classical import ZNE, REM, DD
from qonductor.api import createWorkflow, deploy, workflowResults
from functools import partial

# Define the QAOA circuit and error mitigation techniques
qaoa = QAOA(qubits=10, optimizer='COBYLA')
zne_circuit = ZNE.apply(qaoa, noise_factors=(1, 3, 5)
pre_process = DD.apply(zne_circuit, sequence_type = "XpXm"))
rem_corrected = partial(REM.post_select(counts))
post_process = ZNE.inference(rem_corrected, "LinearFactory") 

#Read deployment configuration file
with open('deployment.yaml', 'r') as file:
    config = yaml.safe_load(file)

# Package a hybrid workflow image
hwi = createWorkflow([pre_process, qaoa, post_process], config)

# Deploy the hybrid image
worfklowID = deploy(hwi)

# Query the workflow execution status and get the results
while workflowStatus(worfklowID) is not 'finished':
    pass
results = workflowResults(worfklowID)
\end{lstlisting}

\myparagraph{\projecttitle{} APIs}
In contrast to the current standard practice, the \projecttitle{} APIs are hardware-agnostic and delegate hybrid resource allocation to the \projecttitle{} leader node. Table \ref{tab:api} shows \projecttitle{}'s APIs, where ``CP'' and ``DP'' stand for control and data plane, respectively. From the user's point of view, there are only four functions. To create workflow images through the workflow manager, clients call \texttt{createWorkflow} with the hybrid code file or a list of classical and quantum functions. To deploy it on \projecttitle{}, clients call \texttt{deploy} with the image ID and the deployment configuration file. Similarly, users call \texttt{invoke} with the image ID to run it.  Lastly, to retrieve the execution results, clients call \texttt{workflowResults}. A toy example using the \projecttitle{} APIs is shown in Listing \ref{code:programming_model}.


\section{\projecttitle{} Resource Estimator}
\label{section:resource_estimator}

The resource estimator generates resource plans that serve two purposes: (1) clients can choose the plan that suits their cost-fidelity tradeoff preferences, and (2) the plans contain meta-information, such as fidelity and execution time estimations, that aid the scheduler in performing the final resource allocation. In this work, we focus on the error mitigation techniques provided by Qiskit \cite{qiskit-error-mitigation, circuit-knitting-toolbox} and Mitiq \cite{mitiq} due to being readily available and easy to integrate.

The resource estimator workflow is shown in Figure \ref{fig:resource-estimator}. First, we apply the error mitigation, which generates one or more circuits (a). Then, we transpile the circuit fragments for template QPU models filtered by the client's preferences (b), to estimate the fidelity (c) and execution time (d) of the mitigated circuits for those models. Finally, we generate resource plans based on the estimated fidelities and aggregate execution times (e). We detail each of the steps below.

\myparagraph{Error mitigation}
The first stage integrates complementary error mitigation techniques in a stacked manner to enhance execution fidelity (\S~\ref{section:background:101}). Specifically, we combine methods that reduce gate, measurement, and decoherence-induced errors at the same time. For instance, REM with Pauli twirling and dynamical decoupling address all major sources of errors. In general, we focus on out-of-the-box techniques provided by Qiskit \cite{qiskit-error-mitigation} and the Mitiq framework \cite{mitiq}, which include: ZNE, PEC, readout error mitigation, dynamic decoupling, Pauli twirling, twirled readout error extinction, probabilistic error amplification, and quasi-probability decomposition implemented as circuit knitting \cite{circuit-knitting-toolbox}. Notably, the core error mitigation techniques (e.g., ZNE, PCE, and circuit knitting) generate multiple circuit instances per input circuit with varying noise characteristics.

\myparagraph{QPU transpilation}
To estimate fidelity and quantum execution time, this step transpiles the generated circuits to template QPUs, after filtering them based on the client's execution preferences. 
Template QPUs adopt the basis gate set and qubit coupling map of a specific QPU model (e.g., IBM Falcon r5.11 \cite{processor-types}), but their calibration data are the average of all available QPUs of that model. Thus, we have as many template QPUs as available QPU models in the system. This coarse-grained approach is scalable since quantum cloud providers typically offer a few models (e.g., up to three in IBM \cite{ibmQuantum}).

\myparagraph{Fidelity and execution time estimation}
To predict both the fidelity and execution time of circuits executed with Qonductor using error mitigation, we employ regression-based prediction models. We first construct a dataset comprising over 7,000 job executions collected from our experiments on the IBM quantum cloud. For either type of estimation, we have to use the type of error mitigation applied as a feature.
For execution time estimation, we use circuit features such as the number of qubits (width), the number of shots, circuit depth, and the number of two-qubit operations. For fidelity estimation, we incorporate additional features, including the qubit topology and error rates of the target QPU. We train and evaluate multiple models through $K$-fold cross-validation, using the $R^2$ score as the primary evaluation metric \cite{gareth2013introduction}. Among the models considered, \textit{Polynomial Regression} yields the highest accuracy, achieving an $R^2$ score of 0.998 for execution time and 0.976 for fidelity prediction.


\begin{figure*} [ht]
    \centering
    \includegraphics[width=\textwidth]{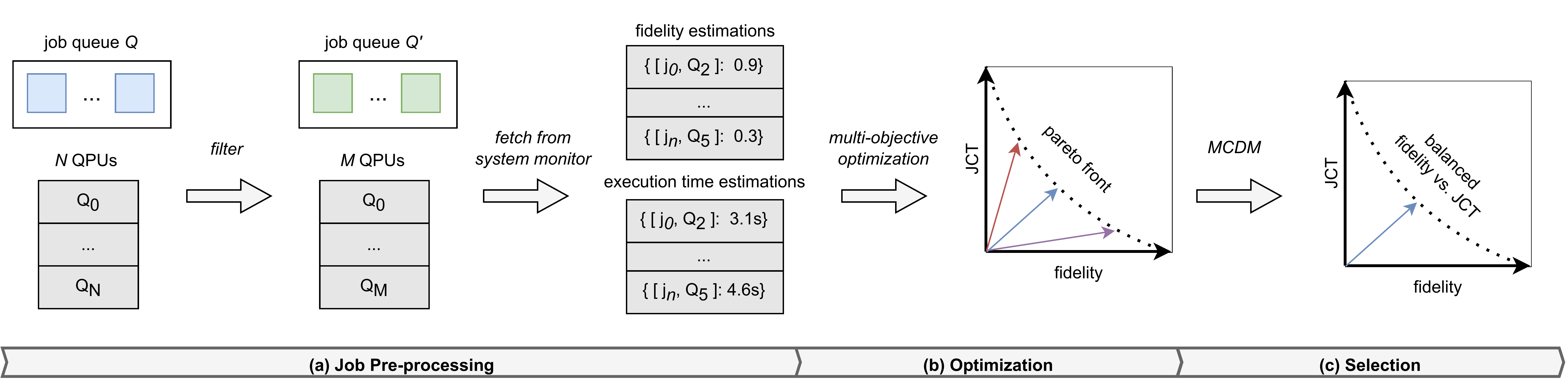}
    \caption{Quantum scheduler workflow (\S~\ref{section:scheduler}). {\em \textbf{(a)} Job pre-processing: the jobs and QPUs are filtered based on the configuration options. Then, the fidelity and execution time estimations are fetched from the system monitor datastore. \textbf{(b)} Multi-objective optimization: we use the NSGA-II genetic algorithm to create a Pareto front of solutions. \textbf{(c)} Selection: we select one of the solutions based on multiple-criteria decision-making (MCDM) that uses pseudo-weights for fidelity and JCT.}}
    \label{fig:scheduler-workflow}
\end{figure*}



\myparagraph{Resource plan generation}
Finally, the resource estimator generates a configurable number of resource plans, by default, three. For this, the estimator stores the estimated fidelity and execution time--which is the sum of quantum and classical execution times-- of the workflow, along with the accelerators used (if applicable) for the error mitigation post-processing step. 

\section{\projecttitle{} Hybrid Scheduler}
\label{section:scheduler}

The \projecttitle{} hybrid scheduler allocates classical and quantum resources to jobs to balance the conflicting objectives of quantum computing, as stated in \S~\ref{section:motivation:characteristics-nisq}. The scheduler supports pluggable policies for heterogeneity and load-aware resource allocation. In \projecttitle{}, we provide two example policies for both types of resources, but mainly focus on quantum job scheduling where prior work is limited.


The scheduling algorithm for classical jobs follows the standard two-stage \textit{filtering-scoring} algorithm of Kubernetes \cite{kubernetes}. For each classical job, the first stage filters the available classical nodes based on the user's configuration file to eliminate incompatible nodes. Based on pluggable scoring policies, the remaining nodes are then scored to find the most suitable nodes. In this work, our default filtering and scoring policies are based on the Kubernetes scheduler \cite{kubernetes-scheduler}, but any heterogeneous-aware resource allocation policy is sufficient. 

The scheduling algorithm for quantum jobs consists of three stages that support configurable policies: (1) the job pre-processing, (2) the optimization, and (3) the selection, as shown in Figure \ref{fig:scheduler-workflow}.

\myparagraph{Job pre-processing}
The first step is to pre-process the jobs to aid the scheduling optimization procedure (Figure \ref{fig:scheduler-workflow} (a)). The scheduler filters the job queue and the QPU list to limit the exploration space and reduce the scheduling overheads. Specifically, it filters out the jobs that cannot run on the cluster given their configuration options (e.g., the system cannot accommodate the client's QPU size requirements). Secondly, the scheduler fetches the fidelity and execution time estimations generated by the resource estimator (\S~\ref{section:resource_estimator}), which are stored in the system monitor. The optimization stage leverages the estimations to generate schedules with fidelity-JCT tradeoffs.

\myparagraph{Optimization}
\label{section:scheduler:multi-objective-optimization}
The optimization stage creates a Pareto front of solutions for the scheduling problem, where the conflicting objectives are fidelity and JCTs (Figure \ref{fig:scheduler-workflow} (b)). In \projecttitle{}, we aim to minimize the mean JCT and maximize the mean fidelity among the scheduled jobs per scheduling cycle, and to do so, (1) we formulate the optimization problem and (2) employ an optimization algorithm to solve it. 

Formally, we formulate the trade-off between fidelity and JCT as follows: $f_1(x)$ is the function capturing the mean JCTs, and $f_2(x)$ is the function capturing the mean error ($1-$mean fidelity), and we aim to minimize both:


\begin{align}
\min \quad & f_1(x) = \frac{1}{N} \sum_{i=1}^N \left(w_{x_i} + \sum_{k=1}^{N} t_{kx_k}[x_i = x_k] \right), \quad 
f_2(x) = \frac{1}{N} \sum_{i=1}^N \left(1 - f_{ix_i} \right) \notag \\
\text{s.t.} \quad & q_{i} - s_{x_i} \leq 0, \quad 1 \leq x_i \leq Q, \quad \forall i = 1,..,N
\end{align}

where $x_i$ is a discrete variable encoding the assignment of job $i$ to QPU $x_i$; $N$ is the number of jobs to be scheduled; $Q$ is the number of available QPUs; $w_{x_i}$ represents the approximate waiting time of the job queue of QPU $x_i$; $t_{kx{_k}}$ is the estimated execution time of job $k$ on QPU $x_k$; $f_{ix_{i}}$ is the estimated fidelity of job $i$ on QPU $x_i$; $q_i$ stands for the maximum number of qubits in job $i$; $s_{x_i}$ is the size of QPU $x_i$. This problem formulation scales independently of the number of QPUs, with a complexity of $\mathcal{O}(N)$ for $N$ jobs to be scheduled. 






The formulated multi-objective optimization problem is Pareto-efficient by definition, and the potential solutions can be explored in parallel; this makes it a good candidate for genetic algorithms. Therefore, we use the \textit{NSGA-II} genetic algorithm \cite{deb2002a} that is robust against local optima and highly parallelizable. We customize the algorithm's genetic operators to thoroughly explore the solution space by initializing the population with random integers, simulating the crossover operation on real values using an exponential probability distribution, and perturbing solutions within a parent’s vicinity using a polynomial probability distribution. Lastly, to avoid prolonged execution, we set maximum thresholds for generations and function evaluations and use a sliding window approach for tolerance termination, evaluating a sequence of generations rather than just the latest one.

\myparagraph{Selection}
\label{section:scheduler:selection}
The solutions of the Pareto front differ in mean fidelity and JCTs of the scheduled jobs, covering the full range between their maximum and minimum values. To select a single solution based on priority on fidelity, JCT, or balanced, we use Multiple-Criteria Decision-Making (MCDM) with pseudo-weights (Figure \ref{fig:scheduler-workflow}, (c)). Calculating pseudo weights involves normalizing the distance to the worst solution concerning each objective, which indicates the solution's location in the objective space. Formally, the pseudo-weight equation is:
\begin{equation}
w_i(x) = \frac{(f_i^{max} - f_i {(x)}) \, /\,  (f_i^{max} - f_i^{min})}{\sum_{m=1}^M (f_m^{max} - f_m (x)) \, /\,  (f_m^{max} - f_m^{min})}
\end{equation}

The pseudo-weight $w_i(x)$ measures the relative importance of the $i$-th objective for solution $x$ within the entire Pareto front, and $f_i^{min}$ and $f_i^{max}$ are the minimum and maximum objective values of objective $i$ over all solutions in the Pareto front. We select the solution $x$ with a vector closest to a desired preference vector $P = (p_1, p_2)$; here $p_1$ is mean fidelity, and $p_2$ is mean JCTs and $p_1 + p_2 = 1$.


\begin{figure*} [t]
    \centering
    \includegraphics[width=\textwidth]{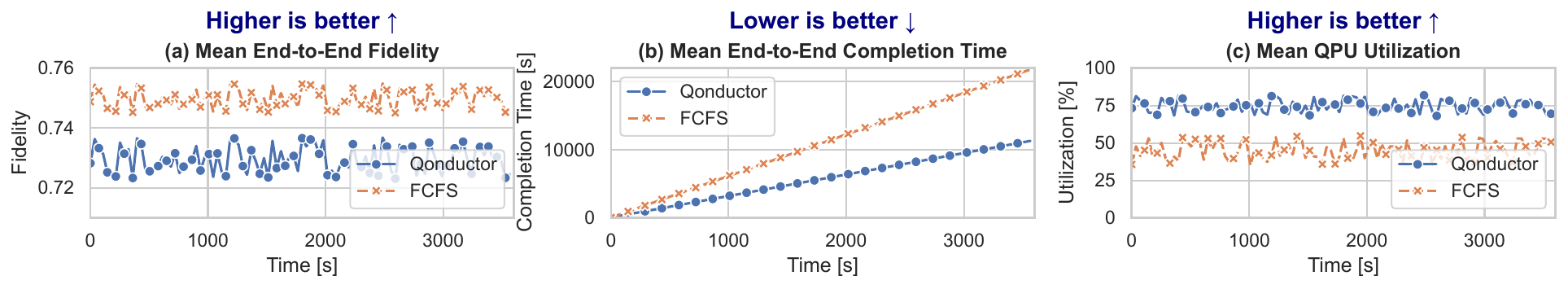}
    \caption{\projecttitle{} end-to-end performance (\S~\ref{section:evaluation:end-to-end}). The experiments run for one simulated hour and 1500 applications/hour. {\em {\bf (a)} Mean end-to-end fidelity: \projecttitle{}'s fidelity is $<3\%$ lower than FCFS. {\bf (b)} Mean end-to-end completion time: \projecttitle{}'s completion times are $\sim48\%$ lower than FCFS. {\bf (c)} Mean QPU utilization: \projecttitle{}'s utilization is $\sim66\%$ higher than FCFS.}}
    \label{fig:end-to-end-performance}
\end{figure*}

\myparagraph{Scheduling triggers}
Scheduling is triggered in two ways: (1) job queue size and (2) time-based trigger. In the former case, if the job queue size reaches a specified limit (100 by default), scheduling is invoked. In the latter case, if a pre-defined time interval elapses (120s by default), scheduling is invoked regardless of the queue size. 

\myparagraph{Calibration crossovers}  
If a generated schedule spans a calibration cycle, we automatically re-evaluate and adjust the jobs scheduled to run after the calibration update. Specifically, our scheduler partitions the schedule at the calibration boundary. Then, it invokes the resource estimator to re-compute fidelity and runtime predictions for jobs in the post-calibration window using the latest calibration data and then reassigns or delays those jobs as necessary. 

\myparagraph{Priority access} In our current implementation, \projecttitle{} does not inherently support provider reservations, as such reservations tend to exacerbate QPU load imbalances by encouraging users to repeatedly select the highest-fidelity QPUs. Instead, when deployed in a cloud environment that implements reservations, Qonductor treats the reserved QPUs as temporarily offline, effectively removing them from the available resource pool during the reservation period. 

\section{Evaluation}
\label{section:evaluation}


\subsection{Evaluation Setup}
\label{section:evaluation:evaluation-setup}

\myparagraph{Implementation}
We implement \projecttitle{} on top of Kubernetes \cite{kubernetes} in Python v.3.11 \cite{python} and Go v.1.21 \cite{go-lang}. In the resource estimator (\S~\ref{section:resource_estimator}), we use the sci-kit-learn v.1.4.0 \cite{scikit-learn} library for estimating fidelity and quantum execution times. For the optimization and MCDM scheduler stages (\S~\ref{section:scheduler}), we employ the pymoo v.0.6.1 \cite{pymoo} framework. Lastly, as our quantum cloud provider, we select IBM Quantum \cite{ibmQuantum} due to its open-access model.

\myparagraph{Experimental setup} 
We conduct two types of experiments: (1) real QPU runs to collect the dataset of the resource estimator (\S~\ref{section:resource_estimator}), i.e., the job execution times and fidelities, and (2) classical simulations of the hybrid cloud.
For (1), we utilize the IBM Quantum open access plan \cite{ibmQuantum} and run jobs on all freely available QPUs. For (2), we run on AMD EPYC 7713P 64-Core servers with 0.5 TB of RAM and use Qiskit's FakeBackends for noisy simulations.

\myparagraph{Benchmarks} We use the MQT Benchmark library \cite{quetschlich2023mqtbench} to generate over 70,000 benchmark circuits, 2 to 130 qubits in size. The library covers all standard quantum algorithms, including VQE \cite{peruzzo2014variational}, Grover's \cite{grover1996fast}, Shor's \cite{shor1999polynomial} algorithms, QAOA \cite{farhi2014quantum}, and Quantum Fourier Transform (QFT) \cite{weinstein2001implementation}.

\myparagraph{Metrics}
For evaluating \projecttitle{}'s performance, we use the following metrics: (1) \textbf{(Job) Completion Time}: Time a job/application requires to complete (\projecttitle{} processing + waiting + executing). (2) \textbf{Fidelity}: We use \textit{Hellinger fidelity} as a measure of the quality of the execution on noisy QPUs \cite{hellinger1909neue, fidelity-qiskit}. Fidelity ranges in $[0,1]$ and higher is better. (3) \textbf{Execution Time}: Time the job runs on a classical or quantum resource, excluding processing and waiting times. 

    \label{fig:scheduler-load-balance}

\begin{figure*} [t]
    \centering
    \includegraphics[width=\textwidth]{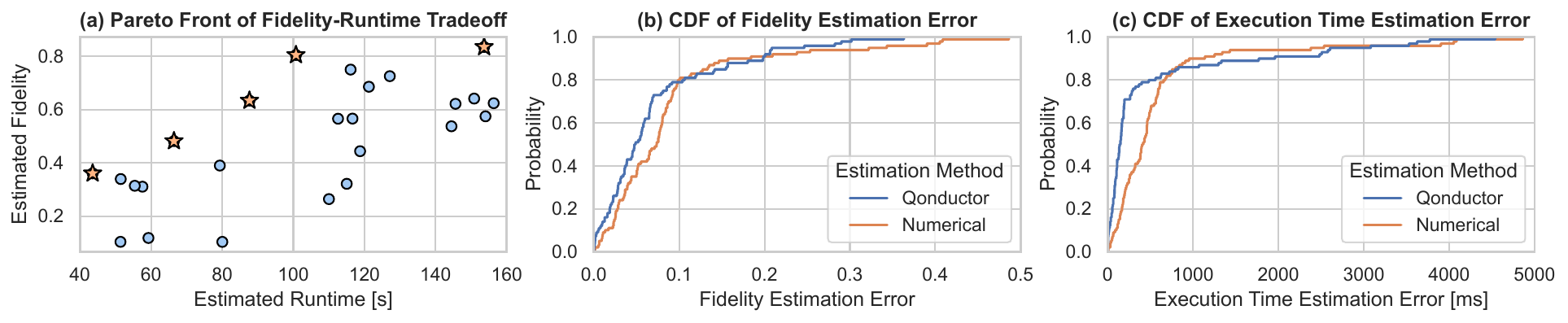}
    \caption{Resource estimator's performance (\S~\ref{section:evaluation:resource_estimator}). {\em {\bf (a)} Pareto front (star points) of the tradeoff between estimated fidelity and hybrid application runtime. {\bf (b)} CDF of the fidelity estimation error. $\sim75\%$ of estimations have an error of less than $0.1$. {\bf (c)} CDF of the execution time estimation error, in milliseconds. $80\%$ of estimations have an error of less than 500ms.}}
    \label{fig:resource_estimator_performance}
\end{figure*}

\myparagraph{Baselines} 
We use the First-Come-First-Serve (FCFS) scheduling algorithm and different configurations of our system as baselines unless otherwise stated. 
Qoncord \cite{wang2024qoncord} specifically addresses the scheduling challenges inherent to \textit{VQAs}, focusing on dividing training iterations between exploratory and fine-tuning phases. In contrast, \projecttitle{} provides a generalized orchestration framework for a broad spectrum of hybrid quantum-classical applications. As such, a direct empirical comparison between Qoncord and \projecttitle{} would not yield meaningful insights, as they are tailored to address distinct challenges within the quantum computing landscape.

\begin{figure*}[ht]
    \centering
    \includegraphics[width=\textwidth]{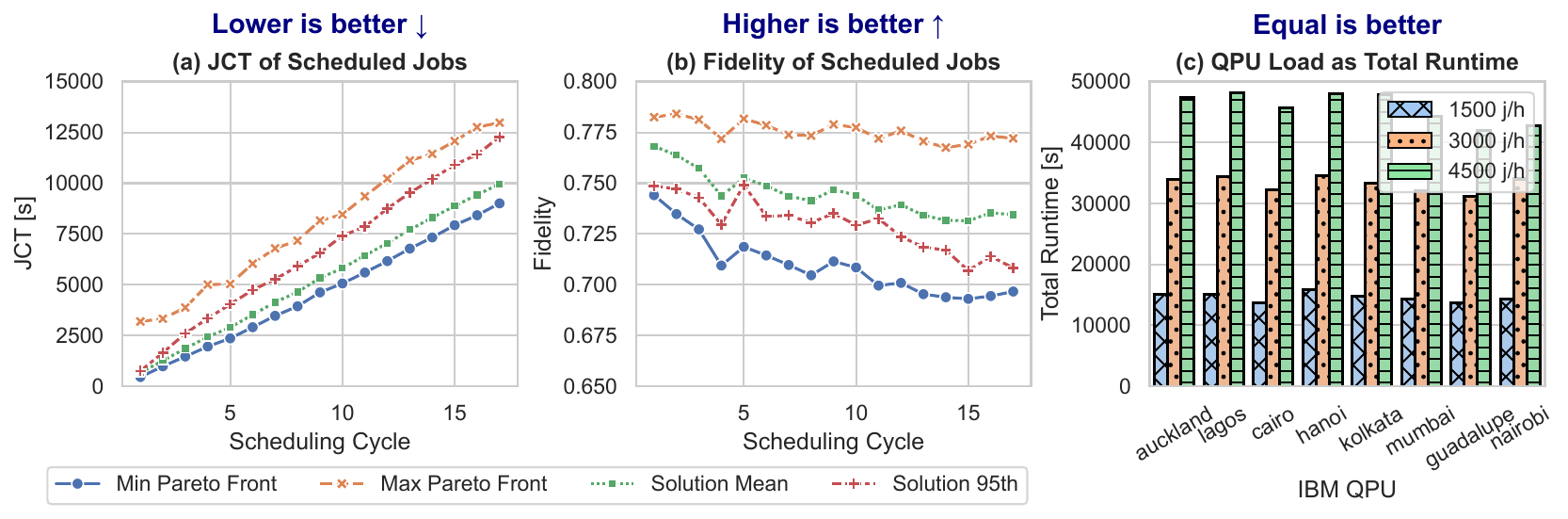}
    \caption{\projecttitle{}'s scheduler performance (\S~\ref{section:evaluation:scheduler-performance}). {\em {\bf (a)} JCTs for scheduled jobs. The chosen solution (mean) incurs 34\% lower and 15.1\% higher JCTs compared to the maximum and minimum Pareto fronts, respectively. {\bf (b)} Fidelities of scheduled jobs. The chosen solution incurs 4\% lower fidelity than the maximum possible. {\bf (c)} QPU load as the total active runtime for increasing workloads. \projecttitle{} distributes the load almost evenly, with a maximum load difference of 15.8\% between QPUs.}}
    \label{fig:scheduler-performance-analysis}
\end{figure*}

\subsection{Quantum Cloud Simulation}
\label{section:quantum_cloud_simulation}

To evaluate \projecttitle{}, we set up a cloud simulation environment replicating the real conditions of the IBM Quantum platform \cite{ibmQuantum}. 

\myparagraph{Dataset collection}
We monitor all available QPUs on the IBM Quantum platform for ten days in November 2023 to gather the QPUs' queue sizes. We then aggregate and analyze the differences in queue sizes for each QPU to measure the job arrival rates. We identify a notable variance in rates across during the day ranging from 1100 to 2050 jobs per hour. The total average of all hours is 1500 jobs per hour and is the baseline system load for our evaluation.

\myparagraph{Load generator}
The load generator creates synthetic workloads that mirror the real-world hybrid application patterns. It generates hybrid applications with random quantum circuits, number of shots, and circuit sizes, following a normal distribution. All applications are transpiled on \projecttitle{}, and a random number of them (50\% on average) use error mitigation, hence utilizing hybrid resources. These applications are then submitted to \projecttitle{} with a fixed frequency, simulating real-world arrival rates. 

\myparagraph{Metrics collection}
We patch Qiskit's FakeBackends with the ability to maintain their own queue of scheduled jobs, job waiting and execution times, and the notion of time flow, reflecting the real-world job flow. After each scheduling cycle, the job manager receives the results and assigns the new jobs to the queues of the chosen backends. 

\subsection{\projecttitle{}'s End-to-End Performance}
\label{section:evaluation:end-to-end}

{\bf RQ1:} \textit{What is the end-to-end performance of \projecttitle{} w.r.t. mean fidelity, completion times, and utilization?} We evaluate \projecttitle{}'s performance by simulating synthetic cloud workloads as stated in \S~\ref{section:quantum_cloud_simulation} and measuring the mean hybrid application fidelity, completion time, and QPU utilization. As a baseline, we use the standard practice in the current quantum cloud, FCFS scheduling. 


Figure \ref{fig:end-to-end-performance} (a) shows the mean fidelity across simulation time. Fluctuations in fidelity are random and depend on (1) the workloads executed at each time point and (2) the application of error mitigation. \projecttitle{}'s mean fidelity is $2-3\%$ lower than that of FCFS since the scheduler selects QPUs with sub-optimal fidelity in favor of completion times. Figure \ref{fig:end-to-end-performance} (b) shows the mean end-to-end completion times across simulation time. \projecttitle{}'s mean completion times are $\sim48\%$ lower than FCFS since \projecttitle{} balances the load across QPUs, reducing the quantum waiting times, while the classical waiting times are practically zero. Notably, both approaches face linearly increasing completion times as a function of time since QPUs are still scarce and, even with load balancing, face long queues. Lastly, Figure \ref{fig:end-to-end-performance} (c) shows the mean QPU utilization across simulation time. Due to load-balancing scheduling and the aforementioned tradeoff exploration between fidelity and completion times, \projecttitle{} achieves $66\%$ higher utilization than FCFS, on average, by distributing the quantum job load across QPUs more evenly.

\myparagraph{RQ1 takeaway} \projecttitle{} achieves $48\%$ lower hybrid application completion times and $66\%$ higher QPU utilization for up to $<3\%$ lower fidelity, compared to FCFS scheduling, on average.


\subsection{Resource Estimator's Performance}
\label{section:evaluation:resource_estimator}

{\bf RQ2:} \textit{How systematically and accurately does the  \projecttitle{} resource estimator explore the fidelity-runtime tradeoff?} We evaluate the resource estimator's performance by visualizing the generated resource plans' fidelity vs runtime and the accuracy of its estimations.

Figure \ref{fig:resource_estimator_performance} (a) shows the fidelity-runtime Pareto front of resource plans, where star points highlight the Pareto-optimal plans. Here, we are using a 20-qubit QAOA max-cut circuit. Each point is a unique resource plan w.r.t. the configuration of error mitigation (e.g. sampling overheads), QPUs used, and classical accelerators for post-processing. Notably, the second-highest fidelity solution incurs 34.6\% lower runtime than the highest, for only 3.6\% lower fidelity. 
\begin{figure*}[t]
    \centering
    \includegraphics[width=\textwidth]{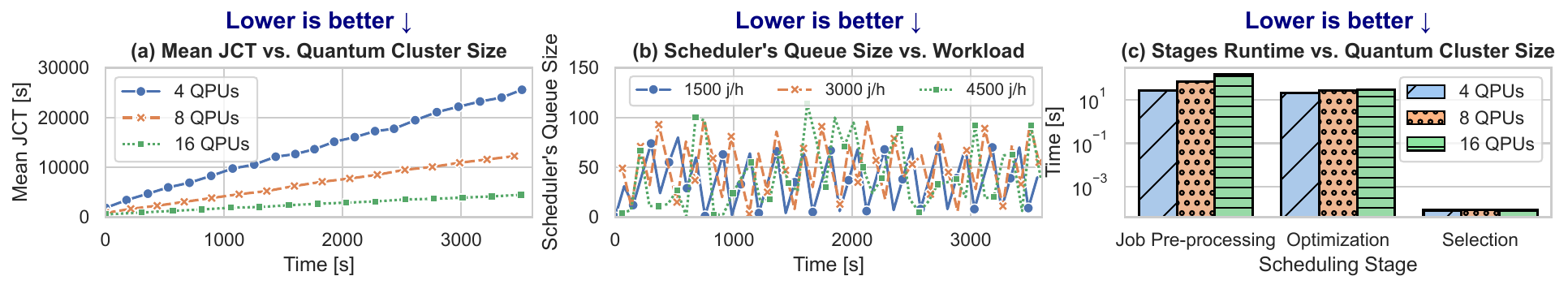}
    \caption{\projecttitle{}'s scheduler scalability analysis (\S~\ref{section:evaluation:scheduler-performance}). {\em (a) Mean JCT as the quantum cluster scales w.r.t. the number of QPUs. As the number of QPUs increases, the mean JCT decreases. (b) \projecttitle{} scheduler scalability w.r.t. system load. The scheduler successfully handles workloads up to 3$\times$ the current IBM load. (c) \projecttitle{} scheduler's stages scalability w.r.t. the cluster size. All stages runtimes are relatively constant as the cluster size increases.}}
    \label{fig:scheduler_scalability_analyis}
\end{figure*}

To measure the resource estimator's estimation accuracy, we plot the absolute difference between the estimated fidelities and quantum execution times with the real, post-execution values, $|est - real|$. The baseline is the numerical approach followed by state-of-the-art work  \cite{wang2024qoncord, zhang2024MECH, mapomatic}, where fidelity and execution times are computed based on the calibration data of the QPU and the operations applied in the circuit, e.g., by traversing the circuit DAG and multiplying the noise errors or summing the gate execution times, respectively. Figure~\ref{fig:resource_estimator_performance}~(b) shows the CDF of the fidelity estimation errors. Our regression model is more accurate than the numerical method by including the effects of error mitigation, although the difference is noticeable only for errors less than $0.1$ fidelity.
Figure \ref{fig:resource_estimator_performance} (c) shows the CDF of quantum execution time estimation error. Here, the regression model outperforms the numerical approach, and 80\% of the estimations are off by less than 500ms. 

\begin{figure*} [t]
    \centering
    \includegraphics[width=0.9\textwidth]{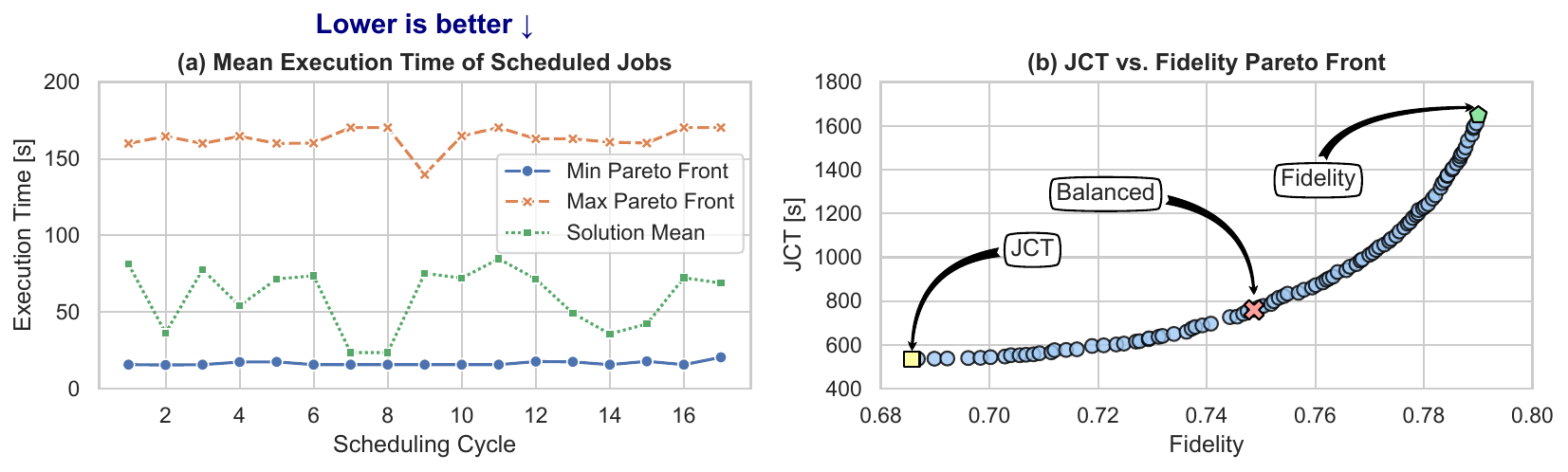}
    \caption{\projecttitle{} scheduler's performance (\S~\ref{section:evaluation:scheduler-performance}). {\em {\bf (a)} Mean execution time of the quantum jobs. The chosen solution achieves 63.4\% lower execution time than the maximum Pareto front.
     {\bf (b)} The scheduler's selection stage (MCDM) selects solutions that match the priorities on conflicting objectives. Balanced: 6\% lower fidelity gives 54\% lower JCT.}}
    \label{fig:mcdm_waiting_time_fidelity}
\end{figure*}

\myparagraph{RQ2 takeaway} Our resource estimator accurately estimates execution fidelity and runtime with minimal error in 80\% of cases and generates resource plans with Pareto-optimal fidelity-runtime tradeoffs.


\subsection{\projecttitle{} Scheduler's Performance}
\label{section:evaluation:scheduler-performance}

{\bf RQ3:} \textit{How well does the \projecttitle{} scheduler balance quantum job fidelity and JCTs, and the load across QPUs?} We use our simulation environment to evaluate the balance between quantum job fidelity and JCTs, the load difference across QPUs, and the mean quantum job execution time.


Figures \ref{fig:scheduler-performance-analysis} (a) and (b) show the minimum and maximum values of the Pareto front for each scheduling cycle, and our chosen solution. Here, the workload is 1500 jobs/hour and we use equal weights between fidelity and JCTs. 
The chosen solutions consistently gravitate towards the minimum Pareto front for JCT. Specifically, the mean and 95th percentile JCTs are 34\% and 17.4\% lower compared to the maximum, respectively. The mean and 95th percentile fidelities are only 4\% and 6\% lower than the maximum, respectively. 

To evaluate load balancing across QPUs, we track the total execution time allocated to each QPU over one hour. The resulting distribution for the eight simulated QPUs is presented in Figure~\ref{fig:scheduler-performance-analysis}~(c), where the load distribution across QPUs is nearly uniform. The maximum load difference between any two QPUs is 15.8\% in the workload case of 1500 jobs/hour.

Lastly, Figure \ref{fig:mcdm_waiting_time_fidelity} (a) shows the mean execution time of the scheduled quantum jobs. The minimum and maximum Pareto fronts are the lower and upper bounds on the mean execution time and act as a proxy to even QPU utilization. The chosen solution achieves 63.4\% lower execution time compared to the maximum Pareto front.

\textbf{RQ3 takeaway.} The \projecttitle{} scheduler successfully manages the tradeoff between fidelity and JCT. The chosen solutions incur 34\% lower JCT for a $4\%$ drop in fidelity. Also, the scheduler balances the load across all available QPUs with minimal load difference ($<16\%$). 



{\bf RQ4:} \textit{How systematically does the \projecttitle{} scheduler create and explore the Pareto front of solutions?} To evaluate this, we generate a synthetic workload of 100 randomly generated quantum jobs and visualize the Pareto front of the generated schedules.

Figure \ref{fig:mcdm_waiting_time_fidelity} (b) shows three different priorities on objectives: JCT, fidelity, and balanced.
By prioritizing JCT over fidelity, the scheduler chooses the solution with the lowest mean JCT, i.e., 67\% lower JCT than priority on fidelity. Inversely, the scheduler chooses the solution with the highest mean fidelity, i.e., 16\% higher than the case of prioritizing JCT. Lastly, assigning equal weights selects a balanced solution, where 6\% lower fidelity leads to 54\% lower JCT.

\myparagraph{RQ4 takeaway} The \projecttitle{} scheduler systematically explores the tradeoff between fidelity and JCTs. Notably, users can experience 54\% lower JCTs for 6\% lower fidelity, on average.

{\bf RQ5:} \textit{How well does the \projecttitle{} scheduler scale with the cluster size (number of QPUs) and the workload (jobs per hour)?} We measure mean JCT improvement with increasing QPU cluster size, the scheduler's pending queue size as the workload increases, and the internal scheduling stages' runtimes with increasing QPU cluster size.

Figure \ref{fig:scheduler_scalability_analyis} (a) shows the mean JCT as the QPU cluster size increases from 4 to 16 QPUs. \projecttitle{} adapts to the growing number of QPUs by utilizing them to evenly distribute the workload. Doubling the system size from 4 to 8 QPUs improves JCTs by 52.8\% and making it four times larger (16 QPUs) improves JCT by 81\% (4.35$\times$ lower).

Figure \ref{fig:scheduler_scalability_analyis} (b) shows the pending job queue size as the workload increases from 1500 j/h to 3000 and 4500 j/h, respectively. The scheduler remains stable even with $3\times$ higher workload than the current, or $\sim$2.2$\times$ the current IBM \textit{peak} workload ($\sim$ 2000 j/h). The oscillation of the three lines reflects the time or window-based triggers of our scheduler. Specifically, each drop in time means that scheduling was invoked, which empties the scheduling queue. Over time, it increases again, until the next scheduling trigger. 

Finally, Figure \ref{fig:scheduler_scalability_analyis} (c) shows the runtime of the three scheduling stages as the QPU cluster size increases. Notably, only the job pre-processing's runtime increases since the fidelity and execution time estimations are performed for more QPUs. We do not compare against an increasing workload (j/h) since the scheduling stages' overheads only scale with the number of QPUs. 

\myparagraph{RQ5 takeaway} The \projecttitle{} scheduler successfully utilizes newly available QPU resources, handles increasing system loads up to 3$\times$ the current IBM system load, and indicates scalable performance for its internal stages.

\section{Related Work}
\label{section:related_work}

\myparagraph{Quantum cloud and serverless}
Research work in quantum cloud computing either analyzes the existing quantum cloud characteristics \cite{ravi2022quantum} or proposes quantum cloud architectures, e.g., serverless \cite{karalekas2020a, sitdikov2023Qiskit, garcia-alonso2022quantum}. However, these approaches do not tackle all three challenges we identify in \S~\ref{section:motivation:characteristics-nisq} in a single system.

\myparagraph{Quantum resource estimation (QRE)}
Research on QRE is limited to predicting the number of physical qubits required to run certain quantum algorithms given a set of assumptions, e.g., the quantum error correction scheme used, if any, and the QPU's calibration data. To our knowledge, Microsoft's Azure Quantum Resource Estimator is the only automated and systematic QRE approach \cite{beverland2022assessingrequirementsscalepractical}. However, it does not account for the fidelity and execution runtime impact of classical resources, in contrast to our approach.

\myparagraph{Quantum Resource Estimation (QRE)}
Azure's Quantum Resource Estimator (QRE) predicts the number of physical qubits needed to run fault-tolerant quantum algorithms under error correction \cite{beverland2022assessingrequirementsscalepractical}. It supports only a limited set of predefined noise models (typically 3–4), making it difficult to extend or adapt to arbitrary hardware. In contrast, Qonductor's estimator predicts both fidelity and execution time for hybrid algorithms on current noisy devices, incorporating the impact of classical resources.

\myparagraph{Quantum job scheduling}
While the field of scheduling has received significant attention in the realm of classical computing, its application to quantum computing remains in its early stages due to the relative infancy of the technology. As a consequence, the area of quantum scheduling is still relatively underdeveloped. 
To the best of our knowledge, quantum job scheduling work is limited to \cite{ravi2021adaptive, salm2022prioritization, kaewpuang2023stochastic, wang2024qoncord, stein2022eqc}. However, these systems face one or more of the following limitations: (1) they implement only single-to-many scheduling, (2) they do not offer fine-grained control over the balance between JCTs and fidelity, (3) they delegate the final scheduling decision to the user, or (4) are limited to VQA algorithms only. In contrast, our many-to-many scheduling policy addresses the needs of both users and cloud providers. It automatically schedules quantum jobs by trading fidelity for JCTs, while balancing the load across QPUs.


\myparagraph{Quantum Resource-sharing}
Existing work on quantum resource sharing focuses almost exclusively on multi-programming \cite{das2019a, liu2021qucloud, niu2023enabling, ohkura2022simultaneous}. Specifically, existing work aims to allocate high-quality regions of the QPU to the bundled programs. Integrating multi-programming is left as future work for our system.

\myparagraph{Classical resource management and scheduling}
Resource allocation and task scheduling on the classical cloud are active areas of research and have been extensively studied. Specifically, a non-exhaustive list of work includes task scheduling \cite{ramezani2013task, guo2012task, panda2015efficient, zhang2018dynamic, li2011cloud}, resource allocation \cite{chang2010optimal, wang2018machine, wang2014multi, dai2016cloud, mireslami2019dynamic, xiao2012dynamic}, and container orchestration \cite{hindman2011mesos, verma2015large, schwarzkopf2013omega, wieder2012orchestrating}. Moreover, there exists work on heterogeneous scheduling \cite{lee2011heterogeneity, wang2014multi, panda2015allocation, panda2015multi} and application-specific scheduling, e.g, for deep learning workloads \cite{xiao2018gandiva, peng2018optimus, narayanan2020heterogeneity, kwon2020nimble}. However, the classical domain does not face the unique challenges of the quantum \S~\ref{section:motivation:characteristics-nisq}. As such, it is not trivial to adapt this work for the quantum cloud. 
\section{Conclusion}

We introduced \projecttitle{}, a cloud orchestrator for developing and deploying hybrid quantum-classical applications on hybrid heterogeneous clouds. To our knowledge, \projecttitle{} is the first holistic approach for hybrid orchestration and improves upon the state-of-the-art in three dimensions: {\bf First,} it exposes hardware-agnostic APIs that abstract the underlying complexity away (\S~\ref{section:programming-model}), {\bf Second,} it offers the first approach to hybrid resource estimation that systemizes tradeoff management in hybrid resource allocation (\S~\ref{section:resource_estimator}), and {\bf Third,} it is the first approach to hybrid scheduling, where for quantum jobs we balance the tradeoff between fidelity and JCTs  (\S~\ref{section:scheduler}). 

\begin{acks}    

We sincerely thank the anonymous reviewers for their feedback. We thank Karl Jansen and Stefan Kühn from the Center for Quantum Technology and Applications (CQTA)- Zeuthen for supporting this work by providing access to IBM quantum resources. Funded by the Bavarian State Ministry of Science and the Arts as part of the Munich Quantum Valley (MQV), grant number 6090181.
\end{acks}

\balance
\bibliographystyle{ACM-Reference-Format}
\bibliography{references}

\end{document}